\documentclass[final]{aipproc}

\newif\ifpdf
   \ifx\pdfoutput\undefined
      \pdffalse 
   \else
      \pdfoutput=1 
      \pdftrue
   \fi

\ifpdf
	\usepackage[pdftex=true,pagebackref=true,pdfstartview={FitH},pdfpagemode={None},pdfauthor={Graeme Lufkin},pdftitle={Disk-Planet Interaction: Triggered Formation and Migration}]{hyperref}
\else
	\usepackage{hyperref}
\fi

\layoutstyle{6x9}

\begin{document}

\title{Disk-Planet Interaction: Triggered Formation and Migration}

\author{Graeme Lufkin}{
	address={Department of Astronomy, University of Washington, Box 351580, Seattle, WA  98195-1580},
}

\author{Thomas Quinn}{
	address={Department of Astronomy, University of Washington, Box 351580, Seattle, WA  98195-1580},
}

\author{Fabio Governato}{
	address={Department of Astronomy, University of Washington, Box 351580, Seattle, WA  98195-1580},
}

\begin{abstract}
We present three-dimensional SPH calculations of giant planets embedded in gaseous disks.
Our findings are collected into a map of parameter space, exhibiting four distinct regions: Type I migration, gap formation, triggered formation of more planets, and wholly unstable disks.
For Type I migration of the planet due to secular interactions with the disk material, the migration rate depends linearly on the disk mass, and is independent of the initial planet mass.
For more massive disks, the planet can disturb the disk strongly enough to trigger the collapse of gas into additional giant planets.
When additional planets form, their interaction as point masses dominates the subsequent behavior of the system.
This mechanism allows for the rapid formation of Jupiter-mass and higher planets.
Migration due to interaction with the disk can significantly change the orbits of giant planets in gas disks.
\end{abstract}

\maketitle


\section{Introduction}
The formation and evolution of giant planets are two important, largely unsolved problems in astrophysics.
Observations of extrasolar planets offer an intriguing puzzle of explaining the presence of many giant planets with very small orbits.
Planet migration via interaction with the gas disk the planet is embedded in appears to have the potential to explain these observations.
Such migration has been explored before, both analytically \citep{ward97,tanaka02} and numerically \citep{nelson00,dangelo02}.
Here we present the results of a new numerical study, using an order of magnitude increase in resolution with a particle-based method of simulation.

\section{Technique}
We simulated a disk of gas around a 1~M$_\odot$ star, specifying a radial temperature profile ($T \propto r^{-1}$) and initial density profile ($\Sigma \propto r^{-3/2}$).
This disk initially extends to a radius of 25~AU, and is free to grow.
The total disk mass was one of the parameters we varied, from 0.01 to 0.2~M$_\odot$.
The disk was composed of $10^5$~gas particles that interact via gravity and Smoothed Particle Hydrodynamics~(SPH).
All simulations included the self-gravity of the disk, were three-dimensional, and had free boundary conditions.
A planet particle, interacting via gravity only, was inserted in the disk on a circular orbit.
The initial mass and semi-major axis of the planet were the other parameters we varied, from 0.25 to 2.0~M$_\mathrm{Jup}$ and 5 to 12.5~AU, respectively.
The disk-planet system was evolved for at least 326~years.
Our simulations were performed with {\em Gasoline}, a tree-based N-body code for gravity and gas dynamics \citep{wadsley03}.

\section{Results}
\begin{figure}
\includegraphics[height=0.35\textheight]{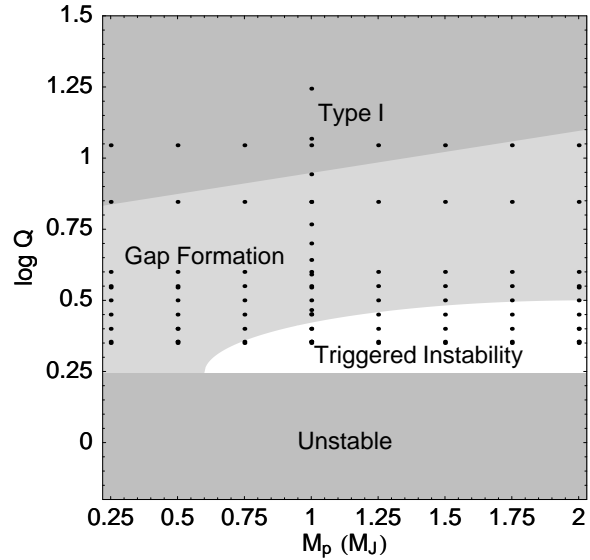}
\caption{A parameter space plot showing the different behaviors observed in a disk-planet system.
The points represent the initial conditions for each simulation.
The labeled regions demarcate the different outcomes we observed after evolving each system for a fixed period of time.
$Q$ is the value of the Toomre stability criteria of the disk at the initial orbit of the planet.}
\label{fig:phase_space}
\end{figure}

Our parameter study of giant planet migration yielded a classification of behaviors, illustrated in Fig.~\ref{fig:phase_space}.
For each point in parameter space (a specific disk mass, initial planet mass, and initial planet orbital radius) we evolved the system for 326~years.
We then classified the state of the system into one of four possible regions: Type I migration, gap formation, triggered formation of additional planets, and wholly unstable disks.

\paragraph{Parameter Map}
In stable disks (the upper region of Fig.~\ref{fig:phase_space}), planets of all masses migrate inward at a roughly constant rate.
In more massive disks, planets can open a gap, with a bias toward more massive planets.
To differentiate between the regions labeled `Type I' and `Gap Formation' we look for the halting of the migration of the embedded planet within the time we simulated.
Thus our definition of a gap is operational; a gap has formed when the density depression along the orbit of the planet has caused the planet to stop migrating.
In yet more massive disks, the density perturbations raised by the inserted planet are large enough to collapse gravitationally, leading to the formation of additional planets.
In this scenario, a disk that is stable in the absence of a ``seed'' planet can fragment to form several additional planets via gravitational instability.
Finally, even more massive disks are unstable even without a seed planet.

The demarcation between regions is based on the amount of time the system has evolved for.
For example, some of the planetary systems in the region labeled `Type I' will in fact open a gap eventually.
So there will be a different parameter space diagram for every age of the systems.
In addition, while we varied three parameters, the diagram is two-dimensional.
We found the disk stability to be the strongest predictor of the eventual outcome.
To accommodate the third parameter, initial orbit size, our diagram displays disk stability at the initial orbit of the planet.
While useful, this contraction of a parameter obscures interesting behavior in the triggered formation of additional planets region.
In this region, close-in planets embedded in less-stable disks did not trigger the formation of new planets, while farther-out planets in more-stable disks did.

\begin{figure}
\includegraphics[height=0.35\textheight]{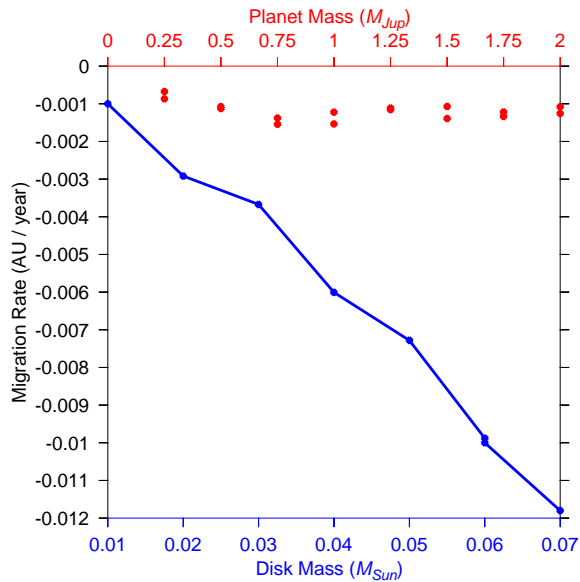}
\caption{The migration rate of a planet embedded in a gas disk, as the planet mass (points) and disk mass (line) is varied.
The migration rate appears independent of the planet mass, and linearly dependent on the disk mass.
The migration was roughly constant over the duration of the simulation (several hundred years).}
\label{fig:migration_rates}
\end{figure}

\paragraph{Migration Rates}
We measured Type I migration rates for several disk-planet systems.
Our findings, shown in Fig.~\ref{fig:migration_rates}, reveal that the migration rate appears to be independent of initial planet mass, and linearly dependent on the disk mass.
This does not agree with the analytic results of \citet{tanaka02}, but is consistent with another numerical investigation, \citet{nelson03b}.

\begin{figure}
\includegraphics[height=0.4\textheight]{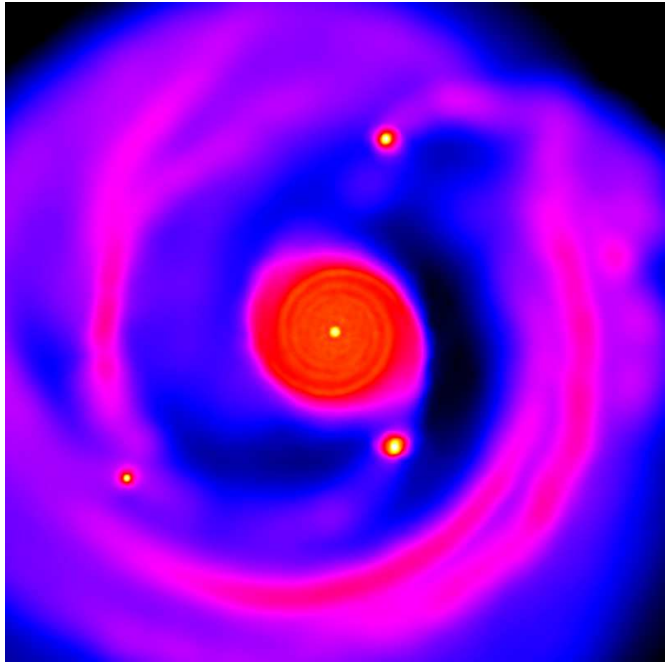}
\caption{A 1.25~M$_\mathrm{Jup}$ planet has triggered the collapse of two additional planets in a 0.1~M$_\odot$ gas disk.
In the absence of the initial planet, this disk will not fragment.
The color represents the logarithmic density of the gas.}
\label{fig:trigger}
\end{figure}

\paragraph{Forming Additional Planets}
The gravitational instability mode of planet formation requires a cool, massive disk \citep{mayer02}.
These conditions can be met locally given a strong density perturbation in the cooler regions of the disk.
The spiral density waves excited by an embedded planet can thus become unstable and collapse into an additional planet.
Our simulations have a fixed temperature profile that falls off with distance from the central star.
Thus we expect that exterior spiral arms would fragment more easily than interior arms.
Fig.~\ref{fig:trigger} is a snapshot of a simulation where this scenario has occurred.
This scenario increases the range of parameter space over which the gravitational instability mode of planet formation is applicable.

Once additional planets have formed, the interaction between the bound objects dominates the orbital evolution of the system.
Therefore, the concept of migration via interaction with the disk is not applicable in this scenario.
Migration via the stochastic process of many-body interaction is still possible.

\section{Conclusion}
This work presents some initial findings on giant planet migration.
A more detailed study of the migration process is necessary to truly understand the interaction between a gas disk and an embedded planet.
Future work will carry out this study, exploring a wider range of parameter space and looking at migration over longer periods of time.
In addition, we have found another region of parameter space where the migration scheme is not applicable, that of triggered formation of additional planets.

\bibliographystyle{aipproc}
\bibliography{gwl}

\end{document}